\documentclass[aps,twocolumn,amssymb,amsmath,amsfonts,superscriptaddress,nofootinbib]{revtex4-1}

\usepackage{hyperref}
\usepackage{breakurl}

\usepackage{graphicx}	% Including figure files
\usepackage{amsmath}	% Advanced maths commands
\usepackage{amssymb}	% Extra maths symbols
\usepackage{float}      % Control for float environments, such as figures

\usepackage{xcolor}
\hypersetup{
    colorlinks,
    linkcolor={red!50!black},
    citecolor={green!50!black},
    urlcolor={blue!80!black}
}

\def\beq{\begin{equation}}\def\eeq{\end{equation}}
\def\bea{\begin{eqnarray}}\def\eea{\end{eqnarray}}

\newfont{\cursive}{pzcmi at 9pt}

\newcommand{\deltecho}{\ensuremath{\Delta t_{\mathrm{echo}}}}
\newcommand{\techo}{\ensuremath{t_{\mathrm{echo}}}}
\newcommand{\ttrunc}{$t_{0}$ trunc.}

\newcommand{\LONGBIB}[2]{#1}

\newcommand{\AUTHAND}{and }
\newcommand{\ARXIVFULL}[1]{\LONGBIB{ [#1]}{}}
\newcommand{\CITELVC}{\LONGBIB{B.~P.~Abbott {\it et al.} [LIGO Scientific \AUTHAND Virgo Collaborations]}{LIGO Scientific \AUTHAND Virgo Collaborations}}
\newcommand{\CITEDOI}[1]{\LONGBIB{doi:#1}{}}

\newcommand{\pkg}[1]{\textsf{#1}}
\newcommand{\pycbcinf}{\pkg{pycbc\_inference }}

\begin{document}

\title{Parameter estimation for black hole echo signals and their statistical significance}

\author{Alex B. Nielsen}
\email{alex.nielsen@aei.mpg.de}
\affiliation{Max-Planck-Institut f\"ur Gravitationsphysik, D-30167 Hannover, Germany}
\affiliation{Leibniz Universit{\"a}t Hannover, D-30167 Hannover, Germany}

\author{Collin D. Capano}
\email{collin.capano@aei.mpg.de}
\affiliation{Max-Planck-Institut f\"ur Gravitationsphysik, D-30167 Hannover, Germany}
\affiliation{Leibniz Universit{\"a}t Hannover, D-30167 Hannover, Germany}

\author{Ofek Birnholtz}
\email{ofek@mail.rit.edu}
\affiliation{Center for Computational Relativity and Gravitation,
  Rochester Institute of Technology,
  170 Lomb Memorial Drive, Rochester, New York 14623, USA}

\author{Julian Westerweck}
\email{julian.westerweck@aei.mpg.de}
\affiliation{Max-Planck-Institut f\"ur Gravitationsphysik, D-30167 Hannover, Germany}
\affiliation{Leibniz Universit{\"a}t Hannover, D-30167 Hannover, Germany}

\begin{abstract}
Searching for black hole echo signals with gravitational waves provides a means of probing the near-horizon regime of these objects.
We demonstrate a pipeline to efficiently search for these signals in gravitational wave data and calculate model selection probabilities between signal and no-signal hypotheses.
As an example of its use we calculate Bayes factors for the Abedi-Dykaar-Afshordi (ADA) model on events in LIGO's first observing run and compare to existing results in the literature.
We discuss the benefits of using a full likelihood exploration over existing search methods that used template banks and calculated p-values.
We use the waveforms of ADA, although the method is easily extendable to other waveforms.
With these waveforms we are able to demonstrate a range of echo amplitudes that is already is ruled out by the data.
\end{abstract}

\maketitle

\section{Introduction}

Black holes are defined by their horizons \cite{Hawking:1973uf}.  Although a
large amount of astrophysical data is compatible with the existence of black
holes \cite{Narayan:2013gca}, a number of theoretical models still predict dark
compact objects without horizons or for which the horizon structure is
significantly modified from classical vacuum general relativity
\cite{Visser:2009pw, Cardoso:2016oxy, Cardoso:2017njb, Cardoso:2017cfl,
Holdom:2016nek}.  These models are typically motivated by quantum effects or
attempts to address issues related to black hole information and evaporation
\cite{Polchinski:2016hrw}.  One possible observational signature of such
structure is that infalling waves would not be entirely absorbed by the horizon
as is generally expected in general relativity, but instead some amount of the
infalling wave would be reflected.

Recent observations of gravitational waves from coalescences of binary black
holes \cite{Abbott:2016blz, Abbott:2016nmj, TheLIGOScientific:2016pea,
Abbott:2017vtc, Abbott:2017gyy, Abbott:2017oio} by the LIGO
\cite{TheLIGOScientific:2014jea} and Virgo \cite{TheVirgo:2014hva} detectors
have allowed for a number of new tests of the near horizon structure of black
holes \cite{TheLIGOScientific:2016src, Cabero:2017avf, Nielsen:2017lpd}.  One
such test involves searching for echo signals that could potentially be caused
by reflective structure forming at or near the location of the black-hole
horizon.  A number of groups have searched for such signals in gravitational
wave data with contrasting conclusions \cite{Abedi:2017, LowSignificance,
Conklin:2017lwb}.  Here we propose a new method to search for these echo
signals that provides an explicit probability for the compatibility of the data
with the echoes hypothesis relative to Gaussian noise.  We demonstrate this
method on the binary black hole events detected during the first observing run
of the Advanced LIGO detectors; these events were the subject of previous
studies \cite{Abedi:2017, LowSignificance, Conklin:2017lwb}.

The general physical picture of echoes is that infalling radiation is reflected
due to some mechanism near the putative horizon location. This radiation is
then partially trapped between the near-horizon structure and the angular
momentum light-ring barrier \cite{Cardoso:2016rao}. Some of the energy is
transmitted away from the system by successive bounces, thereby forming a
series of echoes.  Generic parameters in the physical models are the amount of
wave reflected by the boundary and the effective location where this reflection
occurs. These in turn are related to the amplitude of the reflected echo
signals and the time separation between the successive echoes. Bounds on the
amplitude and time separation of echo signals derived from the data can be
translated into bounds on the reflectivity and location of the near-horizon
structure.

For illustrative purposes here, we focus on the explicit model of
Abedi-Dykaar-Afshordi (ADA) \cite{Abedi:2017},
which has been the subject of discussion in the
literature~\cite{EchoComments, Abedi:2017isz, LowSignificance, Abedi:2018pst}.
However, we note that our methodology can just as well be applied to other,
more detailed models with explicit waveforms, including those recently proposed
in the literature \cite{Mark:2017dnq, Nakano:2017fvh}.
Efforts to search for echo templates using Bayesian model selection
have been developed with LALInference~\cite{lalinf}
in parallel to our own work, and published concurrently with our own \cite{Lo:2018sep}.
Other, model-agnostic searches \cite{Tsang:2018uie}, have also been ongoing, along with different techniques to constrain horizonless objects through their impact on the stochastic background \cite{Barausse:2018vdb}.

The primary result of \cite{Abedi:2017} is a p-value, calculated as the
probability of observing a signal-to-noise ratio (SNR) in noise (assumed to be
free of signal) at least as significant as that observed in the on-source data
that potentially contains the signal.  This by itself does not indicate the
probability that the on-source data contains a signal.
% [Collin] I don't think this sentence is necessary:
%The complement of the
%p-value (one minus the p-value) is the probability that a signal-like feature
%less significant (a statistic less likely) than the one found is obtained in
%noise.
A probability that the data contains a signal can however be obtained
using Bayes' theorem:
\beq \mathrm{P(signal | data ) = \frac{P(data | signal) P(signal)}{P(data)}}. \eeq
It is most convenient to compare this probability to an alternative hypothesis,
for example that the data contains only Gaussian noise:
\beq \mathrm{\frac{P(signal | data )}{P(noise | data )} = \frac{P(data | signal)}{P(data | noise)}\frac{P(signal)}{P(noise)}}. \eeq
In the above, the first factor on the right hand side is the likelihood ratio
and the second factor is the prior odds.  Evaluating the prior odds is
difficult without prior data (and in the case of a signal model that violates
expected physics, might well be a very small factor) but the likelihood ratio
can be calculated by exploring the likelihood function over the model
parameters using a stochastic sampling algorithm, such as a Markov chain Monte
Carlo (MCMC). To obtain a final Bayes factor, the model parameters must be
marginalized over using their respective prior distributions.

The example we consider here is based on the hypothesis of ADA
\cite{Abedi:2017}; we refer the reader to that work for more detail on the
model and the meaning of the various model parameters.
The most important of these parameters are
	the overall amplitude of the echoes relative to the original signal's peak $A$,
	the relative amplitude between successive echoes $\gamma$,	
	and the time separation between successive echoes \deltecho{};
these and the other parameters \techo{} and \ttrunc{} are explained more fully in \cite{Abedi:2017}.
Table~\ref{tab:table1} gives the prior ranges we use for the relevant parameters.
These are adapted for our purposes from the template bank search performed in \cite{Abedi:2017}.
\begin{table}
    \label{tab:table1}
    \begin{tabular}{c|c|c|c} % <-- Alignments: 1st column left, 2nd middle and 3rd right, with vertical lines in between
      \textbf{Echo} & \textbf{Prior} & \textbf{GW150914} & \textbf{Injected} \\
      \textbf {param.}  & \textbf{range} & \textbf{range} & \textbf{value} \\
      \hline
      \deltecho{} & inferred & 0.2825 to 0.3025 s & 0.2925 s\\
      \techo{} & $\deltecho \pm 1\%$ & 0.2795 to 0.3055 s & 0.2925 s \\
      \ttrunc{} & $(-0.1 \mathrm{\; to \;} 0)\deltecho$ & -0.02925 to 0 s & -0.02457 s \\
      $\gamma$ & 0.1 to 0.9 & 0.1 to 0.9 & 0.8 \\
      $A$ & unconstrained & 0.00001 to 0.9 & varying  
    \end{tabular}
     \caption{Table of prior ranges and values used for injection studies. The
     ranges are adopted from \cite{Abedi:2017} and the injected values are
     chosen to lie close to the parameter values found in that work, except for
     $\gamma$ and \ttrunc{} which are chosen to lie within the prior range
     rather than at the boundary.}
\end{table}

In the ADA model the range for \deltecho{} is inferred from the published
parameters of GW150914 \cite{TheLIGOScientific:2016pea}, using $50\%$ ranges,
and assuming Gaussian distributions. The Kerr metric formula is used for the
light travel time between the light ring and a perfectly reflecting surface.
This surface is assumed to be at a proper distance one Planck length along
Boyer-Lindquist time slices from the Kerr metric event horizon.  The parameter
$\gamma$ was chosen to reflect the physical expectation that the amplitude of
successive echoes should decrease due to energy loss through one or both of the
boundaries. We allow the parameter \techo{} to vary independently from
\deltecho{} within $1\%$ of its maximum values, and choose an explicit prior
for the relative amplitude. 

\begin{figure}
  \includegraphics[width=\columnwidth]{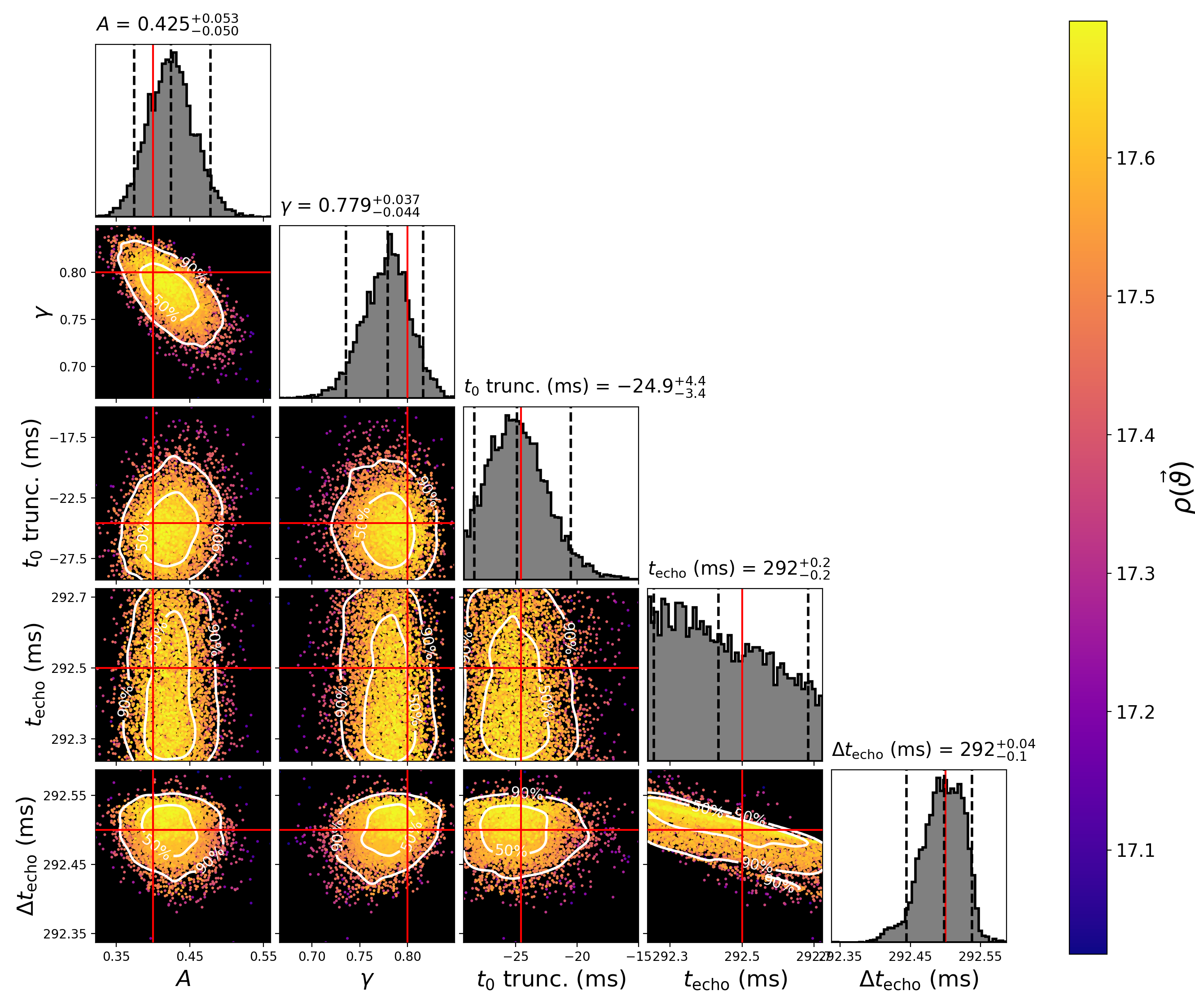}
  \caption{Posterior on the echo parameters for a loud (SNR $\sim 17$)
  simulated signal. The signal has GW150914-like parameters at a fiducial
  distance of $400\,$Mpc. An amplitude factor of 0.4 is used for the echoes.
  Off-diagonal plots show 2D marginal posteriors; the white contours show the
  $50\%$ and $90\%$ credible regions. Each point represents a random draw from
  the posterior, colored by the SNR ($\rho$) at those parameters. The diagonal
  plots show the 1D marginal posteriors, with the median and $90\%$ credible
  intervals indicated by the dashed lines. The reported values are the median
  of the 1D marginal posterior plus/minus the $5/95$ percentiles. We see that the
  injected parameter values, shown by the red lines, are all within the $90\%$
  credible intervals. The log Bayes factor for this signal is $140.57$.}
  \label{fig:dist400_corner}
\end{figure}
\begin{figure}
  \includegraphics[width=\columnwidth]{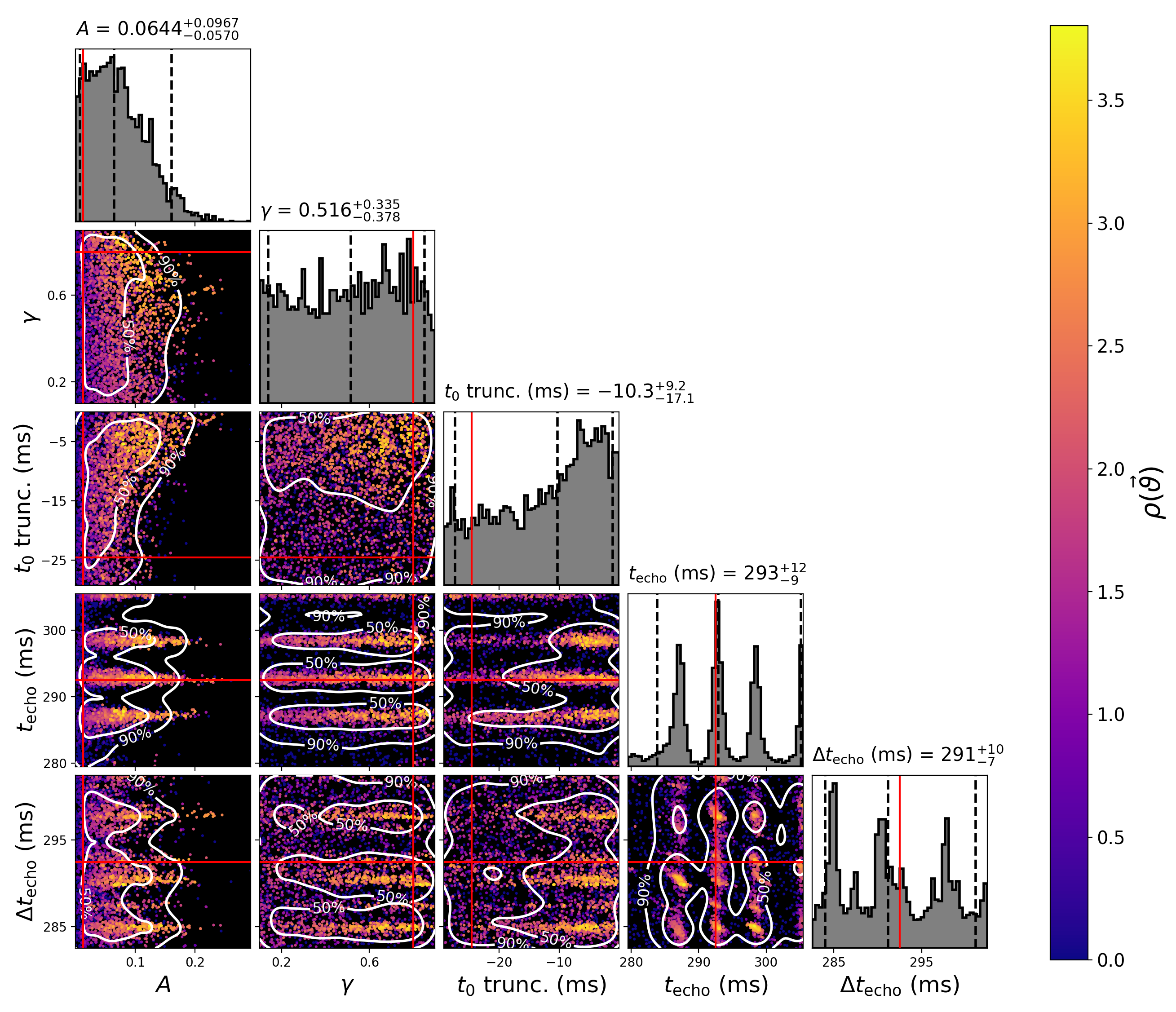}
  \caption{Posterior on the echo parameters for a quiet simulated signal. The
  signal has GW150914-like parameters at a fiducial distance of $400\,$Mpc. An
  amplitude factor of 0.0125 is used for the echoes.  Again, the injected
  values are shown by the red lines, while points are colored by the SNR at
  that point in the parameter space. The log Bayes factor for this injection is
  -1.55, thus indicating what we would expect when the signal is not
  distinguishable from noise. The prior ranges are largely saturated and lines
  appear in \techo{}.}
  \label{fig:dist3200_corner}
\end{figure}

Since the value of the amplitude will have a direct influence on the signal
strength, and hence the signal likelihood, its prior range is of central
importance to our results. In the template bank search of \cite{Abedi:2017} a
prior for the amplitude is not explicitly given. Instead, it is maximised over
the template bank. To replicate as closely as possible the method of
\cite{Abedi:2017} we choose a flat amplitude prior from $10^{-5}$ to $0.9$.
This ensures we are sensitive to relatively quiet amplitude signals, although
not arbitrarily quiet, and implements the reasonable assumption that the first
echo should not be louder than the main signal.

For simplicity we choose to fix the number of echoes to 30. In principle this
could be allowed to vary, but for values of $\gamma$ less than $0.9$, 30 echoes
capture the main part of the signal that influences the SNR.  In testing, we
found that varying this number did not change the results substantially.

To establish that our method can correctly identify echo signals in the data,
we first test it on simulated echo signals with a variety of different
amplitudes.  These simulations are added to real detector data, which is made
available by the Gravitational Wave Open Science Center (GWOSC) \cite{LOSC,
Vallisneri:2014vxa}, $100$ seconds after GW150914.  The $100$-second delay
makes it unlikely that the data at that time is contaminated by a real
astrophysical signal \cite{Nielsen:2018bhc}.  We then apply our method directly
to the three binary black hole events in O1: GW150914, LVT151012 and GW151226.
Finally, we show how these results can be used to place bounds on the
reflectivity of structure that has formed a given distance from the location of
the would-be horizon.

\begin{figure}
  \includegraphics[width=\columnwidth]{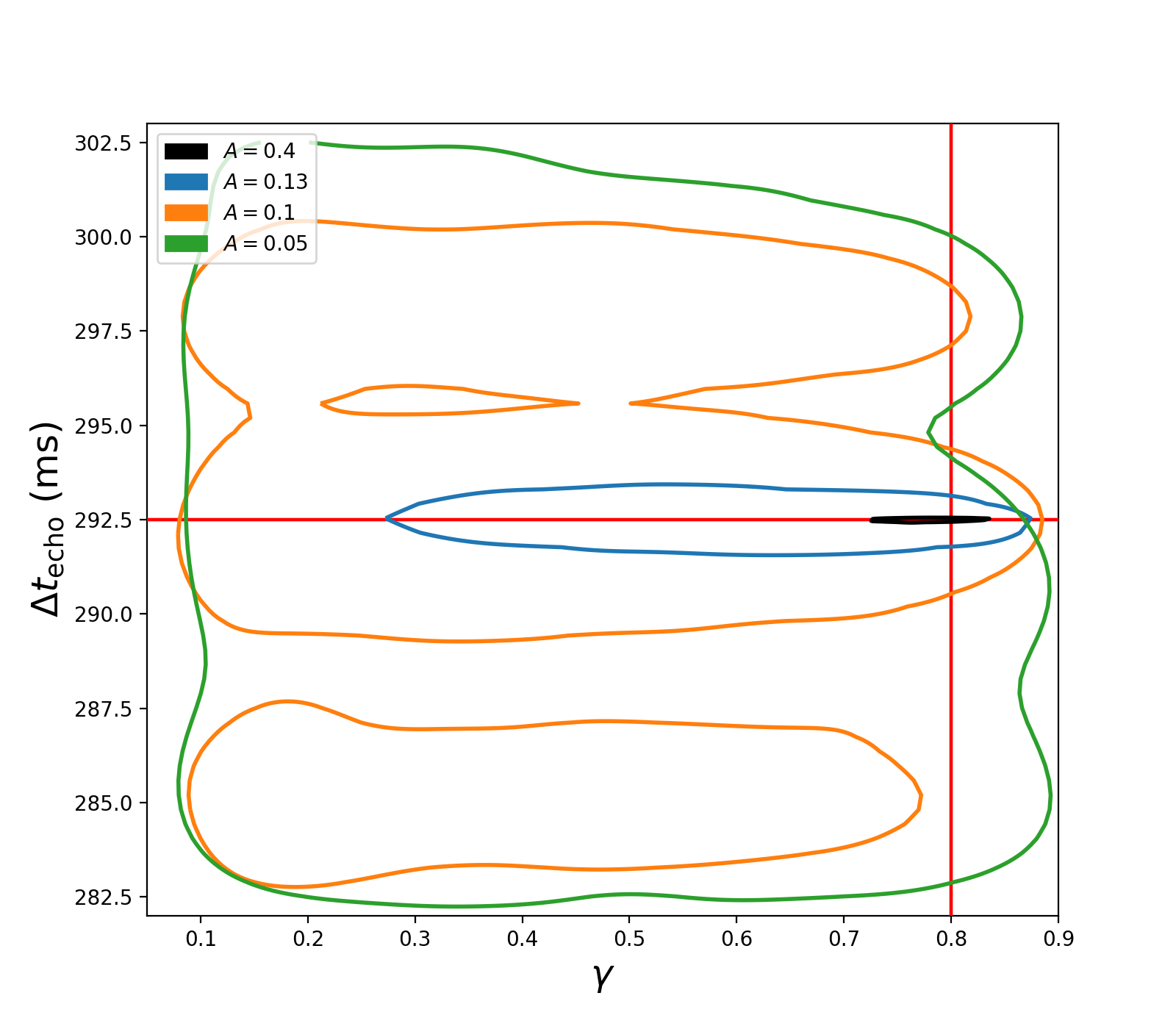}
  \caption{The $90\%$ credible regions of the 2D marginal posteriors of
  $\deltecho$ and $\gamma$ for GW150914-like simulated signals. Shown are a
  range of echo amplitudes (relative to the peak amplitude of the original
  signal) $A$. The injected values are given by the horizontal and vertical red
  lines. For small values of $A$, the $90\%$ contour covers most of the prior
  range, whereas for larger amplitudes the contours narrow down onto the
  injected values.}
  \label{fig:inj_rec_gamma_amplitude}
\end{figure}

\section{Methodology and analysis pipeline}

The pipeline we use is based on \pycbcinf \cite{Biwer:2018osg}. It employs a
parallel-tempered MCMC algorithm, \pkg{emcee\_pt}~\cite{ForemanMackey:2012ig,
Vousden:2016}, to sample the likelihood function for a hypothesis based on the
existence of a signal in the data. The likelihood function is chosen to be
compatible with the assumption that the underlying noise is Gaussian with a
given power spectral density. Once the likelihood has been mapped, the
marginalization over the model parameters is performed using thermodynamic
integration to obtain a probability for the hypothesis given the data. Although
it is known that LIGO data is not Gaussian over long periods of time, over
shorter periods it is approximately Gaussian \cite{Nielsen:2018bhc}. To account
for the non-Gaussianities without a model hypothesis for them, it is possible
to sample the Gaussian Bayes factor over many realisations of the true detector
noise. 

In the results presented here we used 100 Markov chains to sample the
likelihood. We require that each chain run for at least five auto-correlation
lengths (ACL) beyond 1000 iterations of the sampler. The ACL is measured by
averaging parameter samples over all chains, then taking the maximum ACL over
all parameters. For the thermodynamic integration of the likelihood function,
care has to be taken that it is sufficiently sampled both near its peak, but
also at lower values of the likelihood. In tests we found that using 16
different temperatures, each placed by inspection, was sufficient to guarantee
a consistent value of the Bayes factor. Convergence of this result was checked
by running with double the number of temperatures and ensuring that the results
were consistent. The posterior distributions are constructed from the coldest
temperature chain.

\section{Injections based on GW150914}
\label{sec:injections}

To test our method we choose to examine simulated echo signals based on
GW150914.  This is, to date, the loudest binary black hole signal that has been
observed via gravitational waves, and should play a central role in constraints
derived from the data.  Following ADA for simplicity, we choose to fix the base
inspiral-merger-ringdown (IMR) waveform to be echoed for both injections and
for the search templates. The parameters for these base IMR waveforms are given
in the appendix and are obtained from the maximum likelihood results of
\cite{Biwer:2018osg}. The waveforms are constructed
using the phenomenological IMR waveform family IMRPhenomPv2
\cite{Khan:2015jqa,Hannam:2013oca} which is freely available as part of
LALSuite \cite{LALSuite}.
These IMR signals are then used to produce echo signals
with echo parameters given in Table \ref{tab:table1}.  The simulated echo
injections are added linearly at varying amplitudes to real detector noise
(chosen to be 100 seconds after GW150914, far enough away to be uncontaminated
by echo signals or any pre-merger signal). We then attempt to recover them with
our analysis pipeline.  Example results are shown in Figs
\ref{fig:dist400_corner} and \ref{fig:dist3200_corner}.
\begin{figure}
  \includegraphics[width=\columnwidth]{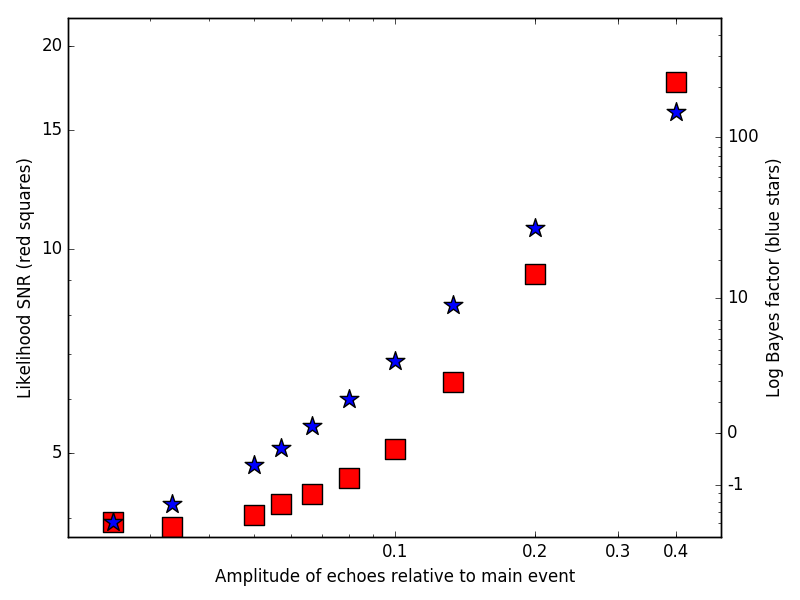}
  \caption{Values of the maximum likelihood SNR and log Bayes factors for
  GW150914-based injections with amplitudes from 0.025 to 0.4 at a distance of
  400Mpc. A linear fit is possible through the SNR points down to an amplitude
  around 0.1. The log Bayes factor is negative for amplitude values below
  $\sim0.07$ (indicating formal preference for Gaussian noise over the echoes
  hypothesis, although at low absolute values of the log Bayes the data is
  uninformative). For amplitudes larger than 0.08, the log Bayes factor is
  greater than 1, indicating positive preference for echoes (the injected
  signal) over Gaussian noise, by the nomenclature of \cite{Kass}.}
  \label{fig:inj_snr_logbf}
\end{figure}

Figure \ref{fig:dist400_corner} shows a very loud injection with a relative
amplitude of 0.4 and a maximum likelihood SNR of $\sim 17.7$. The log Bayes
factor for this injection is 140.57, showing a strong preference for the echoes
hypothesis over the pure Gaussian noise hypothesis. In this case the echo
parameters are well recovered, with the injected values lying within the 90\%
credible intervals of the marginalised one-dimensional posterior distributions.

Figure \ref{fig:dist3200_corner} shows a much quieter injection with a relative
amplitude of $0.0125$ and a maximum likelihood SNR of only $3.8$. The log Bayes
factor for this injection is $-1.55$ showing a preference for the pure Gaussian
noise hypothesis. In this case most echo parameters are not well recovered and
their posterior distributions are close to the original prior distributions.

Figure \ref{fig:inj_rec_gamma_amplitude} shows the recovery of $\gamma$ and
$\deltecho$ for a range of different injection amplitudes. As the amplitude is
increased, the recovered value is increasingly constrained to the injected
value.

The recovery of signals with different amplitudes is shown in
Fig.~\ref{fig:inj_snr_logbf}. This figure can be compared with Fig.~4 of
\cite{LowSignificance}, which shows the recovery of amplitudes relative to the
injected amplitudes. In that work it was found that below a certain injection
strength, the recovered echo amplitude was no longer reliable using the
template bank method. Our results here are consistent with that finding. Here
we find that below an amplitude of $\sim0.1$ the recovered maximum likelihood
SNR no longer falls off linearly and flattens out to an approximately constant
value of $\sim4$. At amplitudes below $\sim0.07$ the log Bayes factor becomes
negative.

\section{Events in the first observing run}
\label{sec:events}

\begin{figure}
  \includegraphics[width=\columnwidth]{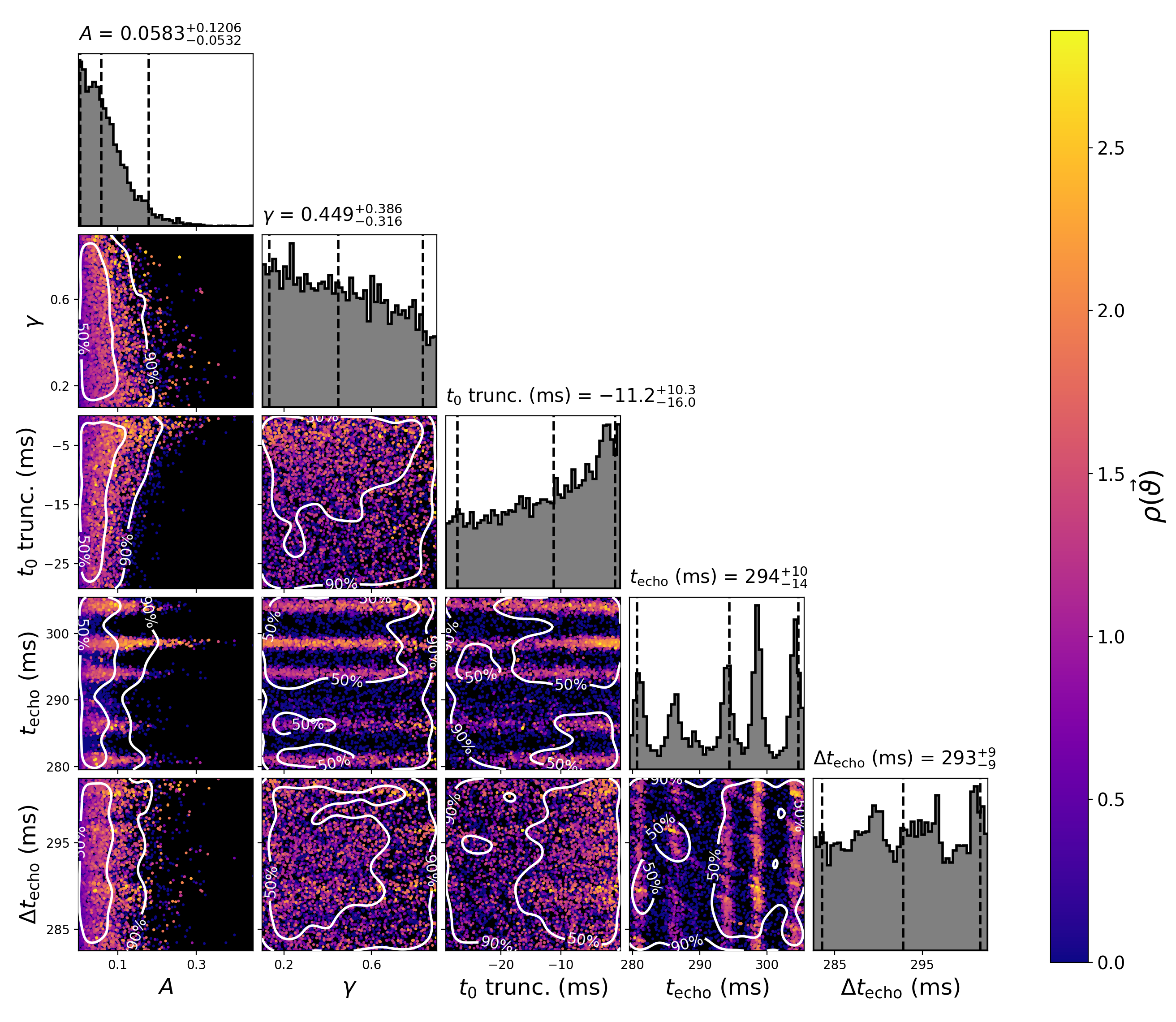}
  \caption{Corner plot for ADA echoes templates in data just after the merger of GW150914. The log Bayes factor for this data is $-1.81$, indicating a preference the Gaussian noise hypothesis over the Echoes hypothesis. Lines are visible in the t\_echo subplots, but the SNR associated with these is still not high. These lines are also seen in tests of the pipeline on simulated Gaussian noise.}
  \label{fig:GW150914_corner}
\end{figure}

The developed pipeline can be run directly on data immediately after the
observed GW events (without injections). We show results for the three events
of the first LIGO observing run in Table \ref{tab:table2}. This shows that
Gaussian noise is favoured over the echoes hypothesis for GW150914 with a log
Bayes factor of $-1.81$. GW150914 is the loudest binary black hole merger yet
detected. A corner plot of the posterior distributions for the echo parameters
for GW150914 is shown in Fig.~\ref{fig:GW150914_corner}. The 90\% credible
interval for the marginalised posterior of the parameter $\gamma$ is almost as
wide as the prior range. The posterior of the amplitude, $A$, prefers lower
values of the amplitude. The posterior for \techo{} shows distinct lines at
certain values of time. These lines are unlikely to be associated with an
astrophysical signal and are also seen in tests on simulated Gaussian noise
with the same pipeline.

As seen in Table~\ref{tab:table2}, both GW151226 and LVT151012 prefer the
echoes hypothesis over Gaussian noise, but only marginally. These two events
have lower amplitudes for the main signal than for GW150914 and thus echoes
signals with the same relative amplitude would have a lower absolute amplitude
relative to the ambient noise \cite{LowSignificance}. The detector noise is
known not to be truly Gaussian for the LIGO detectors
\cite{TheLIGOScientific:2016zmo}. We performed 20 background tests on
off-source data that lies before or after the time of LVT151012 at intervals of
50 seconds. Each of these tests is sufficiently separated in time from the
others that it will not be contaminated by a common signal. In these background
tests, two examples were found with a Bayes factor larger than the result for
LVT151012 shown in Table~\ref{tab:table2}. A total of four intervals returned
Bayes factors that favoured the echo hypothesis over Gaussian noise. Backgrounds for similar (but not identical) echoes hypotheses were also studied in \cite{Lo:2018sep} who found evidence for significant tails in the distribution of Bayes factors in real detector noise versus simulated Gaussian noise.

While it is interesting to speculate whether a signal model could be developed that
postdicts echo signals for certain events, such as LVT151012, but not for
others, such as GW150914, we do not pursue that here. The argument that
LVT151012 should be accepted as a genuine binary black hole merger was given
recently in \cite{Nitz:2018imz}, however we do not feel that the echoes data
for LVT151012 is sufficiently strong to seriously entertain a model where
LVT15012-like events display echoes, but GW150914-like events do not.

The SNR values found for the maximum likelihood templates in
Table~\ref{tab:table2} are comparable, although not identical to those found in
\cite{Abedi:2017} and \cite{LowSignificance}. The values computed here use a
slightly modified echo waveform and the finite template spacing in the template
banks of \cite{Abedi:2017} and \cite{LowSignificance} also causes a minor
difference. The main differences are the different base IMR waveform employed
and the different power spectral density (PSD) used to calculate the matches.
The work of \cite{Abedi:2017} and \cite{LowSignificance} used a PSD directly
from \cite{LOSC} whereas here we have used a PSD computed in \pkg{pycbc}
\cite{Canton:2014ena, Usman:2015kfa} using Welch's method. We estimate the PSD
by taking the median value over 64 8 second-long segments (each overlapped by 4
seconds), centered on the main event.

With the simplistic hypothesis that all three binary black hole events should
show evidence for echo signals in the range of parameters assumed, we can
simply add the log Bayes factor together to obtain an overall log Bayes factor
for this model relative to Gaussian noise of $-1.81 + 1.25 + 0.42 = -0.14$.
This is negative, indicating a preference for Gaussian noise, but not by much.
It is worth noting that this simplistic combination assumes that the values for
the echo parameters can lie anywhere in their prior ranges for any of the three
events. This is slightly different from the hypothesis of \cite{Abedi:2017}
that assumes certain echo parameters should have the same value in all three
events. With a hypothesis that fixes the values of certain echo parameters to
be the same in all cases, it is possible that the overall Bayes factor would be
different from our result. But this issue also raises the question of how these
common parameters should be fixed; a simple maximization of the sum of the
squares of the template SNRs as in \cite{Abedi:2017}, or as a maximization or
marginalization of the likelihood function introduced here. We defer
investigation of these subtle issues to future work.
\begin{table}
    \begin{tabular}{c|c|c|c} % <-- Alignments: 1st column left, 2nd middle and 3rd right, with vertical lines in between
      \textbf{Event} & \textbf{Log Bayes factor} & \textbf{Max SNR}\\
      \hline
      GW150914 & -1.8056 & 2.86 \\
      LVT151012 & 1.2499 & 5.5741 \\
      GW151226 & 0.4186 & 4.07 \\    
    \end{tabular}
     \caption{Table of Bayes factor results. Negative values indicate that the
     Gaussian noise hypothesis is preferred. Positive values indicate that the
     echoes hypothesis is preferred after marginalization over parameters. Log
     Bayes values with magnitude $<1$ are ``not worth  more than a bare
     mention'' in the nomenclature of \cite{Kass}. }
   \label{tab:table2}
\end{table}

\section{Discussion and conclusions}
\label{sec:bounds}

With knowledge of how sensitive our pipeline is from the injection test runs of
Sec.~\ref{sec:injections} we can determine the amplitude of echoes that would
have been detectable had they been present in the data. This allows us to place
a bound on the amplitude of echoes emitted from the events considered here. We remind the reader that bounds from our search only relate to the family of echo waveforms considered here.
These are based on the model proposed in \cite{Abedi:2017} and adopting the prior ranges of Table \ref{tab:table1}.  

As shown in Fig.~\ref{fig:GW150914_corner}, the posterior amplitude recovery has a
90\% confidence interval from $0.0583 + 0.1206 = 0.1789$ to $0.0583-0.0532 =
0.0051$. For this realization of the noise, amplitudes above $0.1789$ are ruled
out at 90\% confidence. This is consistent with the injection studies depicted
in Fig.~\ref{fig:inj_snr_logbf} which show that (for noise at a different time,
100 seconds after the main event) echo signals with amplitudes $\gtrsim0.15$
would have been unambiguously identified in the data.

Echo signals of amplitudes $0.1$ relative to GW150914 would correspond to
approximately 0.1 solar masses of energy being reflected from near the black
horizon \cite{EchoComments}. Although this value of the amplitude is not
conclusively ruled out with the current data, an amplitude as high as 0.2 is
conclusively ruled out by our results.

For numerical simulations of systems similar to GW150914 within general
relativity, it is estimated that approximately 4 solar masses of gravitational
energy flows across the horizon \cite{Gupta:2018znn}. Our constraints here on
the amplitude of echoes within the model of \cite{Abedi:2017} suggest that no
more than $5\%$ of this energy is being reflected by near-horizon structure and
re-emitted as echoes.

We have seen that there is little evidence of ADA echo-like signals in the data of
GW150914. Although there is some evidence of echoes in LVT151012 and GW151226,
as both show positive log Bayes factors, this evidence is not very strong.
Sampling the true detector noise by running over off-source times, shows that
the log Bayes factor found for LVT151012 is not unusual. A number of improved
echo waveform models have been proposed; we defer running with these on further
events to future work.  

\section{Acknowledgments}

O.B. acknowledges the National Science Foundation (NSF) for financial support
from Grant No. PHY-1607520. This work was supported by the Max Planck
Gesellschaft and we thank the Atlas cluster computing team at AEI Hanover. This
research has made use of data, software and/or web tools obtained from the
Gravitational Wave Open Science Center (https://gw-openscience.org), a service
of LIGO Laboratory, the LIGO Scientific Collaboration and the Virgo
Collaboration. LIGO is funded by the U.S. National Science Foundation. Virgo is
funded by the French Centre National de Recherche Scientifique (CNRS), the
Italian Istituto Nazionale della Fisica Nucleare (INFN) and the Dutch Nikhef,
with contributions by Polish and Hungarian institutes.

%\newpage

\appendix*
\section{Fiducial IMR waveform parameters}

We list here the parameters of the base IMR waveforms used to construct the
echo templates both for injections and for the searches. These values are
obtained from the maximum likelihood values of \cite{Biwer:2018osg}.
\begin{table}[H]
    \label{tab:table3}
    \begin{tabular}{c|c|c|c} % <-- Alignments: 1st column left, 2nd middle and 3rd right, with vertical lines in between
      \textbf{Parameter} & \textbf{GW150914} & \textbf{LVT151012} & \textbf{GW151226} \\
      \hline
      mass1 & 39.03 & 22.87 & 18.80 \\
      mass2 & 32.06 & 18.67 & 6.92 \\
      spin1x & -0.87 & 0.12 & 0.44 \\
      spin1y & -0.43 & 0.19 & 0.59 \\
      spin1z & -0.06 & -0.20 & 0.33 \\
      spin2x & -0.11 & 0.018 & 0.00 \\    
      spin2y & -0.03 & -0.019 & -0.017 \\
      spin2z & -0.15 & 0.062 & 0.0033 \\
      distance & 477 & 751 & 315 \\
      ra & 1.57 & 0.65 & 2.23 \\
      dec & -1.27 & 0.069 & 0.98 \\
      tc & 1126259462.42 & 1128678900.46 & 1135136350.66 \\
      polarization & 5.99 & 5.64 & 1.43 \\
      inclination & 2.91 & 2.32 & 0.68 \\
      $\mathrm{coa\_phase}$ & 0.69 & 4.44 & 1.64 \\
      $\mathrm{phase\_shift}$ & -0.92 & -0.91 & 1.86 \\
    \end{tabular}
\end{table}

\end{document}